\documentclass[useAMS,letterpaper,usenatbib]{mn2e}
\pdfoutput=1
\usepackage{graphicx}
\usepackage{array,amssymb,amsmath}
\usepackage[usenames]{color}
\usepackage{mathrsfs}
\usepackage{tikz}
\usepackage{times}
\usepackage{subfigure}
\usepackage{psfrag}

\newcommand{\nc}{\newcommand}

\newcommand{\be}{\begin{equation}}
\newcommand{\ee}{\end{equation}}
\newcommand{\bea}{\begin{eqnarray}}
\newcommand{\eea}{\end{eqnarray}}

\renewcommand\[{\left[}

\newcommand\lsim{\mathrel{\rlap{\lower4pt\hbox{\hskip1pt$\sim$}}
    \raise1pt\hbox{$<$}}}
\newcommand\gsim{\mathrel{\rlap{\lower4pt\hbox{\hskip1pt$\sim$}}
    \raise1pt\hbox{$>$}}}
\def\ee{\end{equation}}
\def\be{\begin{equation}}

\newcommand{\Cp}{{\mathscr{C}}}

\newcommand{\chisq}{\chi^2}
\newcommand{\salt}[1]{{\sc SALT-II}}
\newcommand{\saltorig}[1]{{\sc SALT}}
\newcommand{\sifto}[1]{{\sc SiFTO}}
\newcommand{\mlcs}[1]{{\sc MLCS}}

\nc{\Cm}{\hat{C}}
\nc{\sigint}{\sigma_{\mu}^\text{int}}
\nc{\muth}{\mu^{\text{th}}}
\nc{\cnot}{c_\star}
\nc{\xnot}{x_\star}
\nc{\onesn}{\mathbf{1}_n}
\nc{\ul}[1]{\underline{#1}}
\nc{\diff}{{\mathcal{T}}}
\nc{\lcdm}[1]{$\Lambda$CDM}

\def\reff@jnl#1{{\rm#1\/}}
\def\aj{\reff@jnl{AJ}}                 
\def\araa{\reff@jnl{ARA\&A}}           
\def\apj{\reff@jnl{ApJ}}               
\def\apjl{\reff@jnl{ApJ}}              
\def\apjs{\reff@jnl{ApJS}}             
\def\ao{\reff@jnl{Appl.Optics}}        
\def\apss{\reff@jnl{Ap\&SS}}           
\def\aap{\reff@jnl{A\&A}}              
\def\aapr{\reff@jnl{A\&A~Rev.}}        
\def\aaps{\reff@jnl{A\&AS}}            
\def\azh{\reff@jnl{AZh}}               
\def\baas{\reff@jnl{BAAS}}             
\def\jcap{\reff@jnl{JCAP}}           
\def\jrasc{\reff@jnl{JRASC}}           
\def\memras{\reff@jnl{MmRAS}}          
\def\mnras{\reff@jnl{MNRAS}}           
\def\pra{\reff@jnl{Phys.Rev.A}}        
\def\prb{\reff@jnl{Phys.Rev.B}}        
\def\prc{\reff@jnl{Phys.Rev.C}}        
\def\prd{\reff@jnl{Phys.Rev.D}}        
\def\prl{\reff@jnl{Phys.Rev.Lett}}     
\def\pasp{\reff@jnl{PASP}}             
\def\pasj{\reff@jnl{PASJ}}             
\def\qjras{\reff@jnl{QJRAS}}           
\def\skytel{\reff@jnl{S\&T}}           
\def\solphys{\reff@jnl{Solar~Phys.}}   
\def\sovast{\reff@jnl{Soviet~Ast.}}    
\def\ssr{\reff@jnl{Space~Sci.Rev.}}    
\def\zap{\reff@jnl{ZAp}}               
\def\nat{\reff@jnl{Nature}}            

\title[Comparison of cosmological parameter inference methods]{Comparison of cosmological parameter inference methods applied to supernovae lightcurves fitted with \salt{}}
\author[M.C. March et al.] 
{M.C.~March$^1$\thanks{E-mail: m.march@sussex.ac.uk}, N.V.~Karpenka$^2$, 
F.~Feroz$^3$ 
and M.P.~Hobson$^3$\\ 
$^1$Astronomy Centre, University of Sussex, Brighton BN1 9QH, UK\\
$^2$The Oskar Klein Centre for Cosmoparticle Physics, Department of Physics, Stockholm University, AlbaNova, SE-106 91 Stockholm, Sweden\\
$^3$Astrophysics Group, Cavendish Laboratory, JJ Thomson Avenue, Cambridge CB3 0HE, UK}

\date{Accepted ---. Received ---; in original form 13 July 2012}
\pagerange{\pageref{firstpage}--\pageref{lastpage}}
\pubyear{2012}

\begin{document}
\label{firstpage}
\maketitle
\begin{abstract}
We present a comparison of two methods for cosmological parameter
inference from supernovae Ia lightcurves fitted with the \salt{}
technique. The standard chi-square methodology and the recently
proposed Bayesian hierarchical method (BHM) are each applied to
identical sets of simulations based on the 3-year data release from
the Supernova Legacy Survey (SNLS3), and also data from the Sloan
Digital Sky Survey (SDSS), the Low Redshift sample and the Hubble
Space Telescope (HST), assuming a concordance $\Lambda$CDM
cosmology. For both methods, we find that the recovered values of the
cosmological parameters, and the global nuisance parameters
controlling the stretch and colour corrections to the supernovae
lightcurves, suffer from small biasses. The magnitude of the biasses
is similar in both cases, with the BHM yielding slightly more accurate
results, in particular for cosmological parameters when applied to
just the SNLS3 single survey data sets. Most notably, in this case,
the biasses in the recovered matter density $\Omega_{\rm m,0}$ are in
opposite directions for the two methods. For any given realisation of
the SNLS3-type data, this can result in a $\sim 2 \sigma$ discrepancy
in the estimated value of $\Omega_{\rm m,0}$ between the two methods,
which we find to be the case for real SNLS3 data. As more higher and
lower redshift SNIa samples are included, however, the cosmological
parameter estimates of the two methods converge.
\end{abstract}


\section{Introduction}\label{sec:intro}

It has long been recognised that astronomical standard candles are of
great value in constraining cosmological models. The basic argument is
very straightforward. The luminosity distance $D_{\rm L}$ to an object
of absolute luminosity $L$, from which one measures the flux $F$, is given
by\footnote{For simplicity, we work for the moment in terms of
  bolometric quantities.}
\begin{equation}
D_{\rm L}  \equiv \left(\frac{L}{4\pi F}\right)^{1/2}.
\label{eq:11}
\end{equation}
Thus the distance modulus $\mu = m - M$ between the apparent magnitude $m
\equiv -2.5 \log_{10} F$ of the object and its absolute magnitude $M$ 
is given by
\begin{equation}
\mu = 5 \log_{10} \left(\frac{D_{\rm L}}{\mbox{Mpc}}\right) + 25,
\label{eq:dmod1}
\end{equation}
where the constant offset ensures the usual convention that $\mu=0$ at
$D_{\rm L} = 10$~pc.

In a standard FRW cosmological model containing cold dark matter and dark
energy, defined by the usual cosmological parameters
$\Cp=\{\Omega_{\rm m,0},\Omega_{\rm de,0},H_0,w\}$, the luminosity
distance to an object of redshift $z$ is given by
\begin{equation}
D_{\rm L}(z,\Cp) = \frac{c}{H_0}\frac{(1+z)}{\sqrt{|\Omega_{k,0}|}}
S(\sqrt{|\Omega_{k,0}|}I(z)),
\label{eq:12}
\end{equation}
where 
\begin{equation}
I(z)\!\ \equiv\!\! \int_0^z \!\!\!\frac{d\bar{z}}
{\sqrt{(1+\bar{z})^3\Omega_{\rm m,0} \! + \! 
(1+\bar{z})^{3(1+w)}\Omega_{\rm de,0} 
\! + \! (1+\bar{z})^2 \Omega_{k,0}}},
\end{equation}
in which (neglecting the present-day energy density in radiation)
$\Omega_{k,0} \equiv 1 - \Omega_{{\rm m},0}-\Omega_{\rm de,0}$ and
$S(x)=x$, $\sin x$ or $\sinh x$ for a spatially-flat
($\Omega_{k,0}=0$), closed ($\Omega_{k,0}<0$) or open
($\Omega_{k,0}>0$) universe, respectively.  The special case
$w=-1$ corresponds to a cosmological constant, for which one usually
denotes the present-day density parameter by $\Omega_{\Lambda,0}$.

Thus, by measuring the distance moduli and redshifts of a set of
objects ($i=1,2,\ldots,N$) of known absolute magnitude (standard
candles), and considering the difference
\begin{equation}
\Delta\mu_i = \mu_i^{\rm obs}-\mu(z_i,\Cp)
\label{eq:deltamu}
\end{equation}
between the observed and predicted distance modulus for each object
(often termed Hubble diagram residuals), one can place constraints on
the cosmological parameters $\Cp$. One should note,
however, that in the case where the standard candles share a common,
but unknown, absolute magnitude $M$, this value is exactly degenerate
with the Hubble constant $H_0$, as is clear from (\ref{eq:11}) and
(\ref{eq:12}).

In practice, there are no perfect astronomical standard candles.
Type-Ia supernovae (SNIa), for example, have absolute magnitudes that
vary by about $\pm 0.8$~mag in the $B$-band due to physical
differences in how each supernova is triggered and also due to
absorption by its host galaxy. Nonetheless, SNIa do constitute a set
of `standardizable' candles, since by applying small corrections to
their absolute magnitudes, derived by fitting multi-wavelengths
observations of their lightcurves, one can reduce the scatter
considerably, to around $\pm 0.15$~mag in the $B$-band.  In essence,
SNIa with broader lightcurves and slower decline rates are
intrinsically brighter than those with narrower lightcurves and fast
decline rates (\citealt{Phillips:1993ng,HamuyPhillips1996}). 

SNIa lightcurve fitting techniques fall into two categories: those
that give a direct estimate of the distance modulus, such as the Multi
Colour Light Curve Shape (\mlcs~) method \citep{JhaRiess2007} and
those that give estimates of the supernova apparent magnitude,
lightcurve shape and colour, which can be translated into distance
modulus via supernovae global parameters that must be inferred
simultaneously with the cosmological parameters. This latter category
of techniques includes lightcurve fitting methods, such as the
Spectral Adaptive Lightcurve Template method, (\saltorig{})
\citep{GuyAstier2005,AstierGuy2006}, \salt{} \citep{GuySullivan2010}
and \sifto{} \citep{ConleySullivan2008}. It is this latter category of
methods with which this paper is concerned, and in particular the
\salt{} methodology. Another important difference between the two
categories of lightcurve fitter is that the former infers the SNIa
distance moduli directly, which are then used to infer the
cosmological parameters, whereas the latter divides the process into
two steps: first the lightcurves are fitted to obtain SNIa light curve
parameters such as $\hat{m}^*_{B,i}, \hat{c}_i, \hat{x}_{1,i}$ which
are then used to infer cosmological parameters simultaneously with the
SNIa global parameters $\alpha,\beta,M_0$, in a second step. This
therefore provides an opportunity to use the products of the first
step of the \salt{} analysis, namely the stretch, colour and absolute
B-band magnitude $(\hat{m}^*_{B,i}, \hat{c}_i, \hat{x}_{1,i})$ as the
inputs to alternative methods for inferring cosmological parameters.

In this paper, we take advantage of this opportunity and perform a
comparison of cosmological inference methods using supernovae
lightcurves fitted with \salt{}. In particular, we compare the
standard $\chi^2$-method which is widely used in the analysis of SNIa,
and the Bayesian hierarchical method\footnote{A copy of the BHM code
  is available from the corresponding author on request.} (BHM), which
was recently proposed by \cite{MarchTrotta2011}. For varying
implementations of the standard $\chi^2$-method, see for example
\cite{AstierGuy2006,Kowalski2008Improved,Kessler2009Firstyear,Amanullah2010Spectra,GuySullivan2010,ConleyGuy2011,MarrinerBernstein2011}. The
comparison is performed by applying both methods to sets of realistic
simulated SN data based on the real 3-year data release from the
Supernova Legacy Survey (SNLS3) \citep{ConleyGuy2011,GuySullivan2010},
together with a compilation of various other samples at lower and
higher redshift suggested and supplied by the SNLS3 team.  We also
apply both inference methods to the real SNLS3 single survey data set
to compare the cosmological parameter inferences obtained from the two
approaches.

The outline of this paper is as follows. In
Section~\ref{sec:cosmoinf}, we give a short summary of \salt{}
lightcurve fits, followed by a brief description of the standard
$\chi^2$-method and BHM for inferring cosmological parameters from
SNIa lightcurves fitted with \salt{}. In Section~\ref{sec:realdata},
we describe the real SNLS3 data set along with the real SDSS, HST and
LowZ samples on which our simulations are based, and then discuss our
simulation process in Section~\ref{sec:simdata}. The statistical
comparison of the $\chi^2$-method and BHM, as applied to our simulated
data sets, is described in Section~\ref{sec:simresults}, and the
results of applying the two methods to the real data are presented in
Section~\ref{sec:realresults}. We conclude in Section~\ref{sec:conc}.

Finally, we note that this paper may be considered as complementary to
our companion paper (Karpenka et al. 2012), in which we use an
extension of the BHM to constrain the properties of dark matter haloes
of foreground galaxies along the lines-of-sight to (a subset of) the
SNIa in the SNLS3 catalogue, assuming a fixed background cosmology.

\section{Cosmological inference methods}
\label{sec:cosmoinf}

Cosmological parameter inference takes place after the selection cuts,
lightcurve fitting and Malmquist correction stages of the analysis
process have been implemented (see Section~\ref{sec:realdata}).  The
most widely used method for cosmological parameter inference from
\salt{} fitted lightcurves is the basic minimization of the chi-square
statistic, although there are a few differences in the way in which
this method is implemented, as outlined below. More recently, \cite{MarchTrotta2011}
proposed a Bayesian hierarchical method (BHM), which provides a robust
statistical framework for the full propagation of systematic
uncertainties to the final inferences.  We give a brief outline of
these two alternative approaches below, but note that in our
subsequent analyses the SNIa data input to the two methods are the
same, having had the same selection cuts, fits and corrections made.

For each selected SNIa, in addition to an estimate $\hat{z}$ of its
redshift and an associated uncertainty $\sigma_z$, derived from
observations of its host galaxy, we take as our basic data the output
from the \salt{} lightcurve fitting method, which produces the
best-fit values: $\hat{m}_{B}^\ast$, the rest frame $B$-band apparent
magnitude of the supernovae at maximum luminosity; $\hat{x}_1$, the
stretch parameter related to the width of the fitted light curve; and
$\hat{c}$, the colour excess in the $B$-band at maximum
luminosity. These are supplemented by the covariance matrix of the
uncertainties in the estimated lightcurve parameters, namely
\begin{equation}
\widehat{C} = 
\left(
\begin{array}{ccc}
\sigma^2_{m^\ast_B} & \sigma_{m^\ast_B,x_1} & \sigma_{m^\ast_B,c} \\[2mm]
\sigma_{m^\ast_B,x_1} & \sigma^2_{x_1} & \sigma_{x_1,c}\\[2mm]
\sigma_{m^\ast_B,c} & \sigma_{x_1,c} &  \sigma^2_{c} \\
\end{array}
\right).
\label{eq:covmat}
\end{equation}
Therefore, our basic input data for each SN
($i=1,\ldots,N_{\rm SN}$) are
\begin{equation}
D_i \equiv
\{\hat{z}_i,\hat{m}_{B,i}^\ast,\hat{x}_{1,i},\hat{c}_i\},
\label{eqn:inputdata}
\end{equation}
and we assume (as is implicitly the case throughout the SN literature)
that the vector of values
$(\hat{m}_{B,i}^\ast,\hat{x}_{1,i},\hat{c}_i)$ for each SN is
distributed as a multivariate Gaussian about the true values, with
covariance matrix $\widehat{C}_i$. 

\subsection{Standard $\chi^2$-method}

The standard method of cosmological parameter inference used with
outputs from the \salt{} lightcurve fitter is a basic chi-square
minimization technique, see for example
\cite{AstierGuy2006,Kowalski2008Improved,Kessler2009Firstyear,Amanullah2010Spectra,GuySullivan2010,ConleyGuy2011,MarrinerBernstein2011}. A
detailed account of this approach, together with a description of some
statistical issues associated with the methodology, is given in the
preceding references; we therefore present only a brief summary here.

One begins by defining the `observed'
distance modulus $\mu^{\rm obs}_i$ for the $i$th SN as
\begin{equation}
\mu_i^{\rm obs} = \hat{m}_{B,i}^\ast - M + \alpha \hat{x}_{1,i} - \beta\hat{c}_i,
\end{equation}
where $M$ is the (unknown) $B$-band absolute magnitude of the SN,
and $\alpha$, $\beta$ are (unknown) nuisance parameters controlling
the stretch and colour corrections; all three parameters are assumed
to be global, i.e. the same for all SNIa.

One then defines the $\chi^2$ misfit function
\be \label{eq:chisq}
\chisq(\Cp,\alpha,\beta, M,\sigma_\text{int}) = \sum_{i=1}^N 
\frac{[\mu_i^\text{obs}(\alpha,\beta,M)-\mu_i(\Cp)]^2}{\sigma_i^2(\alpha,\beta,\sigma_\text{int})},
\ee
where, for clarity, we have made explicit the functional dependencies
of the various terms on (only) the parameters to be fitted. In this
expression, $\mu_i$ is the predicted distance modulus given by
(\ref{eq:dmod1}) and is a function of SN redshift $z_i$ and the
cosmological parameters $\Cp$, and the total dispersion $\sigma_{i}^2$ 
is the sum of several errors added in quadrature:
\be
\sigma_{i}^2 = \sigma_{z,i}^2 + \sigma_\text{int}^2 
+ \sigma_{\text{fit},i}^2(\alpha,\beta).
\ee
The three components are: (i) the error $\sigma_{z,i}$ in the redshift
measurement owing to uncertainties in the peculiar velocity of the
host galaxy and in the spectroscopic measurements; (ii) the intrinsic
dispersion $\sigma_\text{int}$, which describes the global variation in
the SNIa absolute magnitudes that remain after correction for stretch
and colour; and (ii) the fitting error, which is given by
\be
\sigma_{\text{fit},i}^2 = \bpsi^{\rm t} \Cm_i  \bpsi,
\ee
where the transposed vector $\bpsi^{\rm t} = \left(1, \alpha, -\beta
\right)$ and $\Cm_i $ is the covariance matrix given in
(\ref{eq:covmat}).


Typically, the chi-squared function (\ref{eq:chisq}) is minimized
simultaneously with respect to the cosmological parameters $\Cp$ and
the global SNIa nuisance parameters $\alpha$, $\beta$ and $M$.  There
are, however, a few differences in the way in which this minimisation
is performed, such as which search algorithm is used (MCMC techniques
or grid searches) and the treatment of $M$ (which is degenerate with
$H_0$), namely whether these parameters are marginalised over
analytically or numerically. Once this chi-squared minimisation has
been performed, the value of $\sigma_\text{int}$ is estimated by
adjusting it to obtain $\chi^2/N_{\rm dof} \sim 1$, usually by some
iterative process.


In this work, we take the simple\_cos\_fitter algorithm\footnote{Alex
  Conley's simple\_cos\_fitter code has generously been made available
  at: http://qold.astro.utoronto.ca/conley/simple\_cosfitter\/}
\citep{ConleyGuy2011} as representative of the general class of
chi-square methods, and compare its performance with the BHM, which we
describe below. It should be noted that the chi-square method is an
approximation to the BHM in certain limits (see \citealt{Gull1989} for a
general discussion of this, and \citealt{MarchTrotta2011} for a
discussion of this as applied to the SNIa case.). Hence we expect the
two methods to converge in some limit; detailed studies into this
exact limit have not yet been carried out.

\subsection{Bayesian hierarchical method}
\label{sec:cosmoinf:bhm}

Recently, a more statistically well-motivated method for cosmological
parameter inference from SNIa was put forward in the form of a
Bayesian hierarchical model \citep{MarchTrotta2011}, itself based on the methodology of \citep{Gull1989}, and indeed is a special case of the more general methodology of \citep{Kelly2007}; which provides for full propagation
of systematic uncertainties to the final inferences and also allows
for rigorous model selection.


In essence, this method provides a means for constructing a robust
effective likelihood function that yields the probability of obtaining
the observed data for the $i$th SN (i.e. the parameter values obtained
in the \salt{} lightcurve fits) as a function of the
cosmological parameters and global SNIa nuisance parameters, namely
\begin{equation}
\Pr(\hat{m}_{B,i}^\ast,\hat{x}_{1,i},\hat{c}_i,\hat{z}_i|\Cp,\alpha,\beta,
\sigma_\text{int}),
\label{eqn:likelihood}
\end{equation}
which also depends on the covariance matrix $\widehat{C}_i$ of the
uncertainties on the input data
$(\hat{m}_{B,i}^\ast,\hat{x}_{1,i},\hat{c}_i)$, and the uncertainty
$\sigma_{z,i}$ in the estimated redshift $\hat{z}_i$, all of which are
assumed known. The full likelihood function is given by the product of
the likelihoods (\ref{eqn:likelihood}) for each SN. 

The likelihood for each SN is computed by first introducing the hidden
variables $M_i$, $x_i$, $c_i$ and $z_i$, which are, respectively, the
true (unknown) values of its absolute $B$-band magnitude, stretch and
colour corrections, and redshift. These are then assigned priors,
which themselves contain further nuisance parameters, and all the
parameters introduced in this way are marginalised over to obtain the
likelihood (\ref{eqn:likelihood}). The details of this procedure are
given in Appendix A. By assuming separable Gaussian priors on the
hidden variables and nuisance parameters, one can perform all the
marginalisations analytically, except for two nuisance parameters
$R_x$ and $R_c$, which are also described in Appendix A, that
must be marginalised over numerically.

The full likelihood function is then multiplied by an assumed prior
(see Appendix A) on the unknown parameters to yield their posterior
distribution. This posterior is explored using the MultiNest algorithm
\citep{Feroz2008,Feroz2009}, which implements the nested sampling method \citep{Skilling2004,Skilling2006} adapted for
potentially multimodal distributions.

\section{Real supernovae data}
\label{sec:realdata}

We will perform our comparison of the standard chi-square method and
the BHM by applying them to two classes of data sets. The first of
these is the single survey SNLS3 data alone, i.e.  only SNIa
data which was taken during the first three years of the SNLS3 survey
\citep{GuySullivan2010}.  We take the SNLS3 data as supplied by the
SNLS3 team\footnote{SNLS3 data and associated data sets are available
  from: http://hdl.handle.net/1807/26549} after selection cuts have
been made, the \salt{} lightcurve fitting process has been completed
and the Malmquist correction has been applied.  Our interest in
comparing the performance of the two cosmological parameter inference
methods as applied to this single survey data set is driven by
potential science applications that use a single SNIa survey in
conjunction with other data sets (not SNIa) to investigate various
astrophysical and cosmological phenomena. An example of such an
application is constraining the properties of dark matter haloes of
foreground galaxies along the lines-of-sight to the SNIa using
gravitational lensing, as discussed in our companion paper (Karpenka
et. al. 2012).

%
\begin{figure*}
\centering
\includegraphics[width=0.80\linewidth]{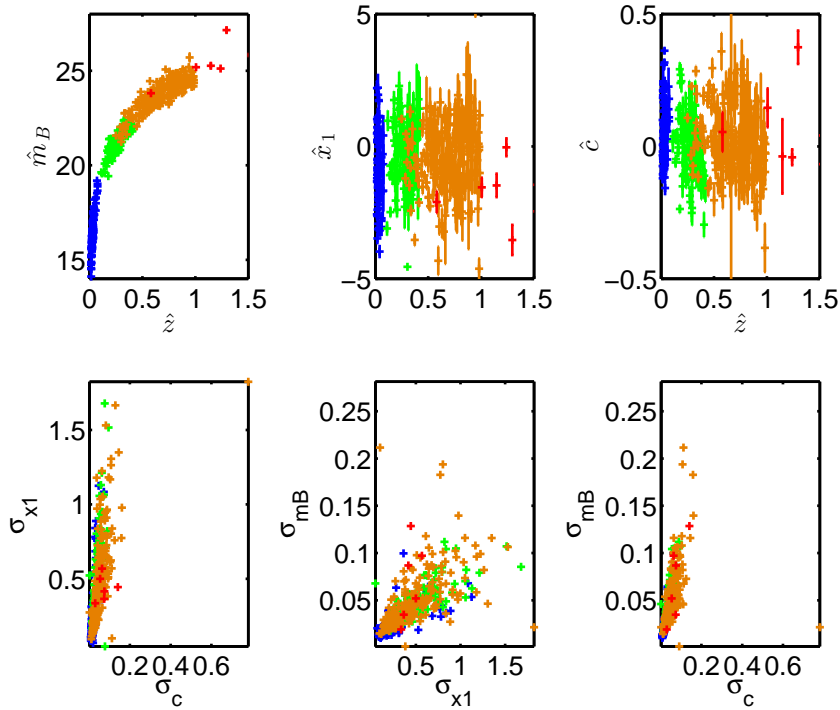} 
\caption{A simulated SNANA data set, comprising SNe from 
Low Z (red); SDSS (green); SNLS3 (gold) and HST (red).}
\label{fig:simSNLS3data}
\end{figure*}

For more general cosmological parameter inference applications, we are
also interested in comparing the performance of the two methods on a
composite `cosmological' data set, since extracting good constraints
on cosmological parameters from SNIa surveys requires the inclusion of
both low-redshift and high-redshift samples.  Since no survey
currently covers both the low- and high-redshift parts of the Hubble
diagram, several data sets are generally considered together to span
the full redshift range.  In this paper we use the compilation
suggested and supplied by the SNLS3 team, which is described in
\cite{GuySullivan2010}. In addition to the SNLS3 data, it is comprised
of: various samples with $z<0.1$; the SDSS survey
\citep{Kessler2009Firstyear, HoltzmanMarriner2008} for $0.1<z<0.4$;
and high-redshift HST data \citep{Riess2007}.  Selection cuts, \salt{}
lightcurve fitting and Malmquist corrections are made by the SNLS3
team and are already implemented in the supplied data files, as
discussed in detail in \cite{PerrettBalam2010} .

\section{Simulated supernovae data}
\label{sec:simdata}

SNIa  photometric data were simulated and fitted using the publicly
available SNANA package\footnote{The SNANA package has generously been made available at:  http://sdssdp62.fnal.gov/sdsssn/SNANA-PUBLIC/} \citep{Kessler2009SNANA}. First, data were
simulated to match closely the SNLS3 data set \citep{GuySullivan2010}
by using the SNLS3 co-added simulation library files (which are
publicly available as part of the SNANA package), a coherent magnitude
smearing of $0.12$, and colour smearing. The colour smearing effect,
or broad-band colour dispersion model, implemented in the data
simulation is the EXPPOL model described by Fig.~8 of
\citep{GuySullivan2010}, and the simulated Malmquist bias is based on
Fig.~14 of \cite{PerrettBalam2010}. In
total 100 similar data sets of SNLS3 style SNIa were simulated.

The SNANA SNIa data simulation is a two-stage process that mimics the
real data collection and analysis process. The first stage is the
simulation of photometric data in accordance with the characteristic
instrument and survey properties of the SNSL3 survey using the SNLS3
simulation library files mentioned above.  The second stage is the
lightcurve fitting process in which the photometric data are fitted to
\salt{} templates to give estimates of the SNIa absolute B-band
magnitude $\hat{m}_B$, lightcurve stretch $\hat{x}_1$ and colour
$\hat{c}$. At this lightcurve fitting stage, basic cuts are made to
discard SNIa with a low signal-to-noise ratio and/or too few observed
epochs in sufficient bands. After the lightcurve fitting stage we make
a redshift dependent magnitude correction for the Malmquist bias; the
correction is taken from a spline interpolation of table 4 in
\cite{PerrettBalam2010}. All selection cuts and Malmquist corrections
take place prior to the cosmology inference step.

In addition to simulating the SNLS3 single survey data sets, we also
simulate combined `cosmological' data sets made by simulating
individual LowZ, SDSS and HST samples which are generated separately
using SNANA and the appropriate simulation templates.  The LowZ sample
SNIa were simulated based on the CFA3-KepplerCam lightcurves
\citep{Hicken2009} ; SDSS sample uses the SDSS 2005 templates
\citep{HoltzmanMarriner2008,Kessler2009Firstyear,Lampeitl2009} and the
HST sample is based on the \cite{Strolger2006} templates. An example
combined simulated data set is shown in Fig.~\ref{fig:simSNLS3data}.

\section{Results from simulated data}
\label{sec:simresults}

In section \ref{sec:simresultslcdm} we analyse both the single survey
SNLS3 data sets, and the combined multi survey `cosmology' data sets
within the framework of the \lcdm~ model in which non-zero curvature
is allowed.  In section \ref{sec:simresultswcdm}, we further analyse
the compiled `cosmology' data sets within the framework of the flat
$w$CDM model in which curvature is fixed at zero, but the dark energy
equation of state, $w$ is permitted to vary.

\subsection{\lcdm~ analysis of SNLS3 and `cosmological' samples}
\label{sec:simresultslcdm}

In the \lcdm~ analysis of simulated SNLS3 data alone, we find that both
methods perform well, considering the difficulties that arise when
attempting to obtain cosmological constraints from a survey that does
not include a low-$z$ sample to anchor the lower end of the Hubble
diagram. 

\begin{figure}
\centering
\includegraphics[width=1.05\linewidth]{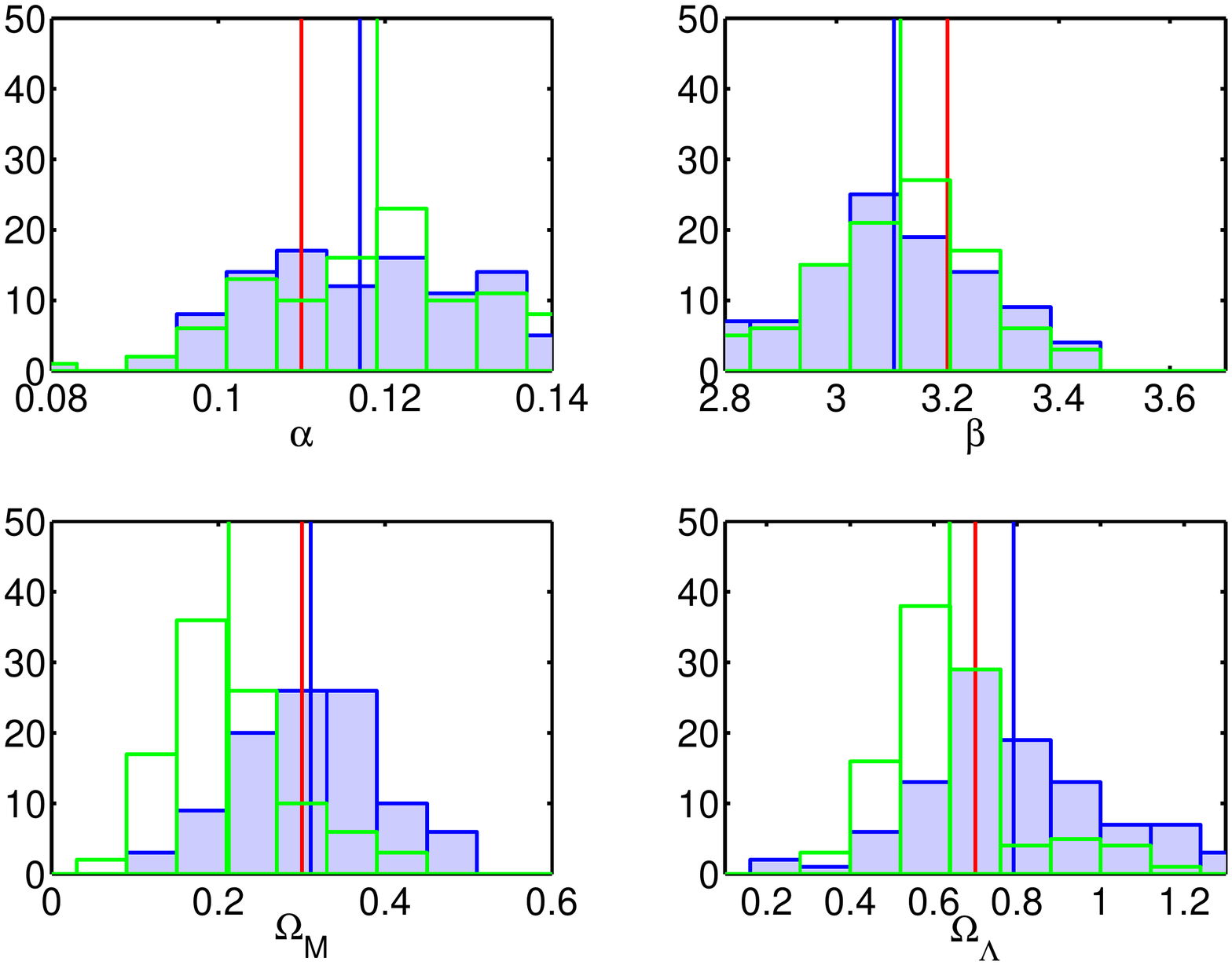}
\includegraphics[width=1.05 \linewidth]{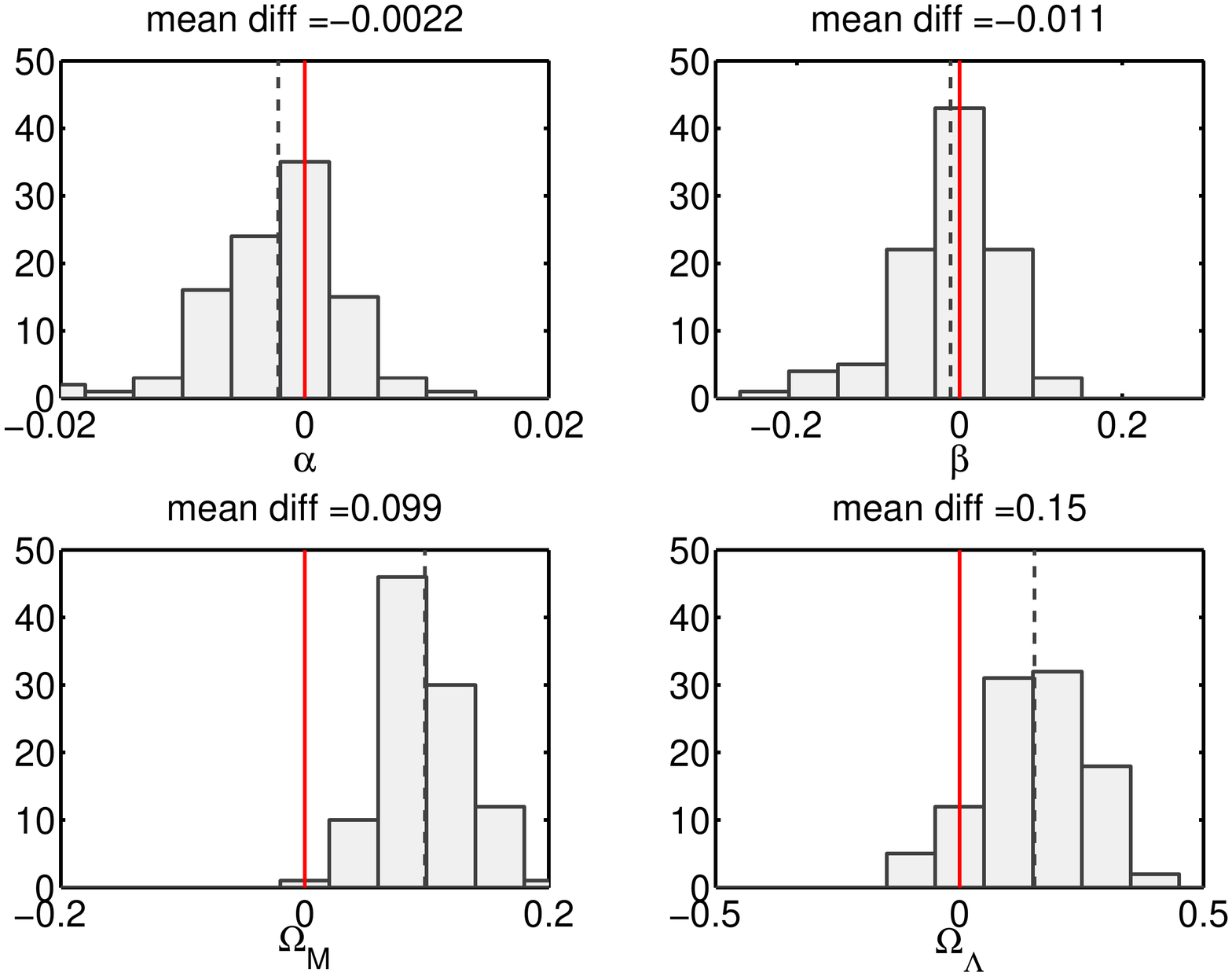}
\includegraphics[width=1.05\linewidth]{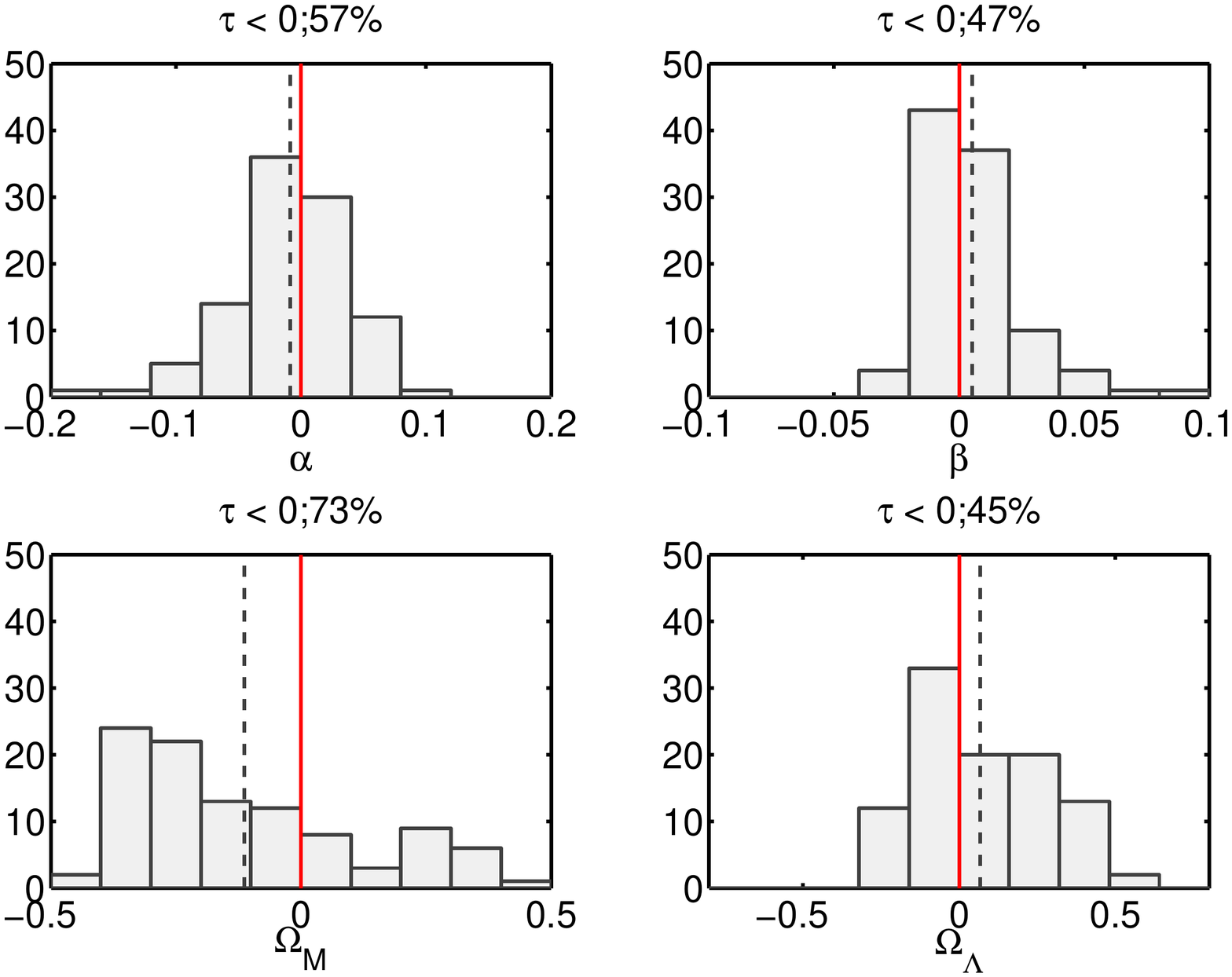} 
\caption{\lcdm~ analysis of simulated SNLS3 data sets.  Top 4 panels:
  sampling distributions derived from 100 sets of simulated data for
  the SNIa global parameters $\alpha,\beta$ and the cosmological
  parameters $\Omega_{\rm m,0},\Omega_{\Lambda,0}$, inferred using the
  BHM (blue histograms) and the chi-square method as implemented by
  the `simple\_cos\_fitter' code (green histogram). Blue and green
  vertical lines show the mean value of the estimator, solid red
  vertical lines show the value of the true (i.e. model input)
  parameter. Middle 4 panels: histograms showing the raw difference in
  estimators between the BHM method and `simple\_cos\_fitter' method,
  given by $\hat{\theta}_{\text{BHM}}-\hat{\theta}_{\chi^2}$. Bottom 4
  panels: the difference in accuracy between the two methods, given by
  $\tau=\theta^{-1}_{\text{true}}
  (|\hat{\theta}_{\text{BHM}}-\theta_{\text{true}}|-|\hat{\theta}_{\chi^2}-\theta_{\text{true}}|)$. Negative
  values of $\tau$ indicate that the BHM estimator is closer to the
  true value than the simple\_cos\_fitter (i.e.$\chi^2$) estimator.
}
\label{fig:lcdm-snls3}
\end{figure}

Fig.~\ref{fig:lcdm-snls3} (top 4 panels) shows the sampling
distributions of the estimators for the parameters of interest for the
BHM (blue histograms) and the $\chi^2$-method (green histograms), as
implemented by the `simple\_cos\_fitter'. We see that both methods
recover similar global SNIa parameters $\alpha,\beta$. More
importantly, both methods recover the cosmological parameters
$\Omega_{\rm m,0}$ and $\Omega_{\Lambda,0}$, but with small biasses
that differ between the methods.

\begin{table}
\centering
\begin{tabular}{lrr|rr}
\hline\hline 
                      &\multicolumn{4}{c}{\lcdm} \\
                      &\multicolumn{2}{c}{SNLS3 only}&\multicolumn{2}{c}{Combined sample} \\ \hline
Parameter             &BHM      & $\chi^2$-method & BHM    & $\chi^2$-method  \\ \hline
$\alpha$              &$0.007$  &$0.009$          &$0.010$&$0.0021$               \\
$\beta$               &$-0.096$ &$-0.085$         &$-0.029$&$-0.075$               \\   
$\Omega_{\rm m,0}$    &$0.011$  &$-0.088$         &$-0.037$&$-0.041$               \\
$\Omega_{\Lambda,0}$  &$0.091$  &$-0.062$         &$0.021 $&$0.0086$               \\             
\hline \hline
\end{tabular}
\caption{Bias on the estimators of the parameters of interest obtained
  from the two cosmological inference methods, for \lcdm~model, as a result of the analysis of 100 sets of SNSL3 only data, and 100 sets of `cosmology' data sets.}
\label{tab:bias_SNLS3lcdm}
\end{table}

The biasses on the recovered parameter values
for both methods are listed in Table~\ref{tab:bias_SNLS3lcdm}.
Perhaps most notable, is that the recovered value of $\Omega_{\rm
  m,0}$ in BHM is biassed slightly high, whereas that for the $\chi^2$
method is biassed somewhat low.  Thus, when the two methods are used
to analyse the same data set, they can give discrepant values for
$\Omega_{\rm m,0}$.  Fig. \ref{fig:lcdm-snls3} (middle 4 panels) shows that
the average discrepancy in $\Omega_{\rm m,0}$ for the single survey
SNLS3 data is $\sim 0.1$ and the maximum discrepancy can be up to
$\sim 0.2$. From the bottom 4 panels of 
Fig.~\ref{fig:lcdm-snls3} we see that in $\sim73\%$ of
trials the BHM provides an estimator for $\Omega_{\rm m,0}$ which is
closer to the true value of $\Omega_{\rm m,0}$ than the estimator
given by the $\chi^2$ method.

Analysis of the SNLS3 survey alone is useful for particular
applications such as when used in conjunction with other non SNIa data
sets as mentioned earlier. However, when the primary aim of the
analysis is cosmological parameter inference, then several different
surveys are analysed together to span an appreciable redshift range
and form a `cosmological' sample. In order to obtain constraints on
the cosmological parameters, we analysed combined `cosmological'
samples described in section \ref{sec:simdata} which span a broader
redshift range.  

\begin{figure}
\centering
\includegraphics[width=1.0 \linewidth]{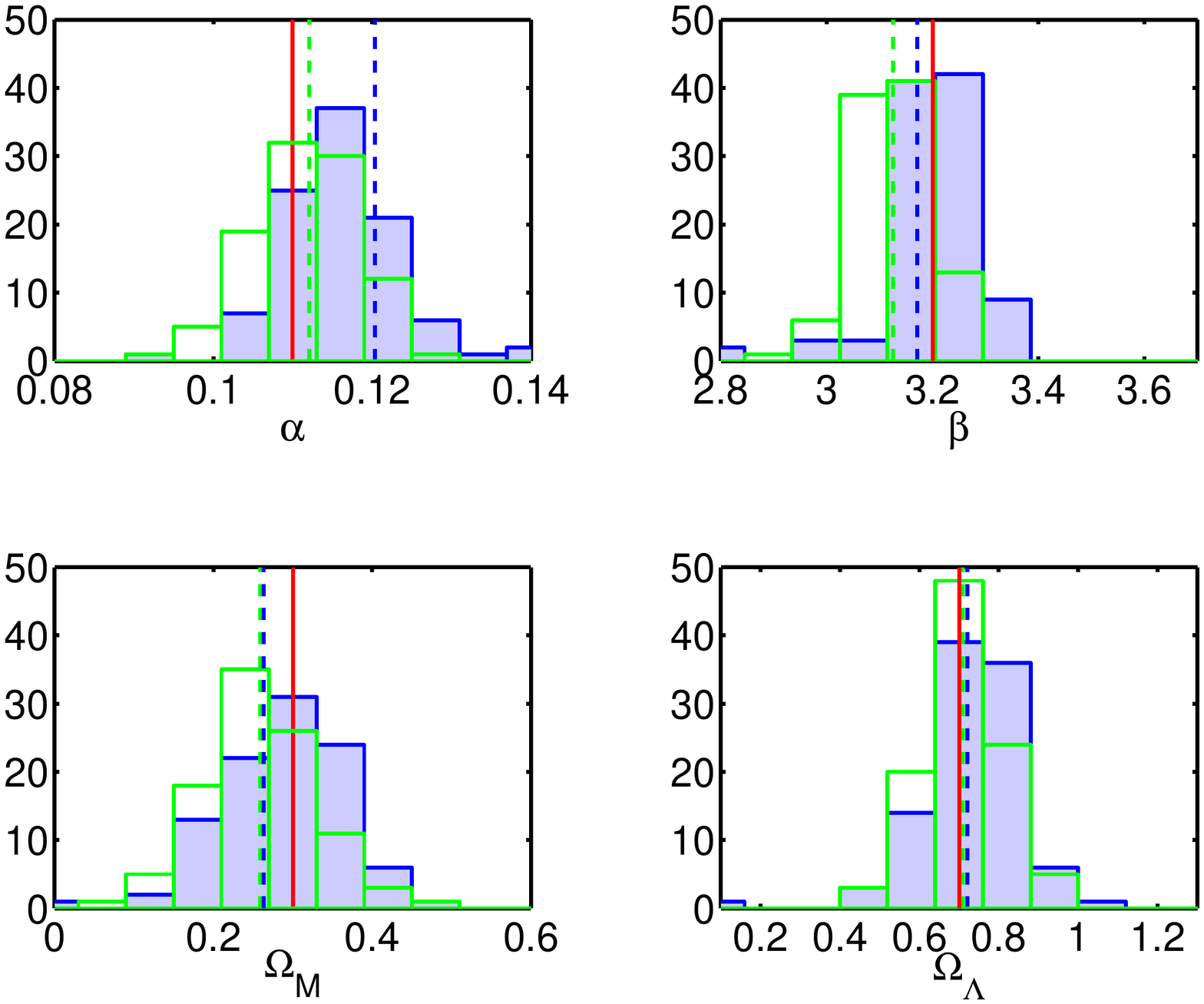}  
\includegraphics[width=1.0 \linewidth]{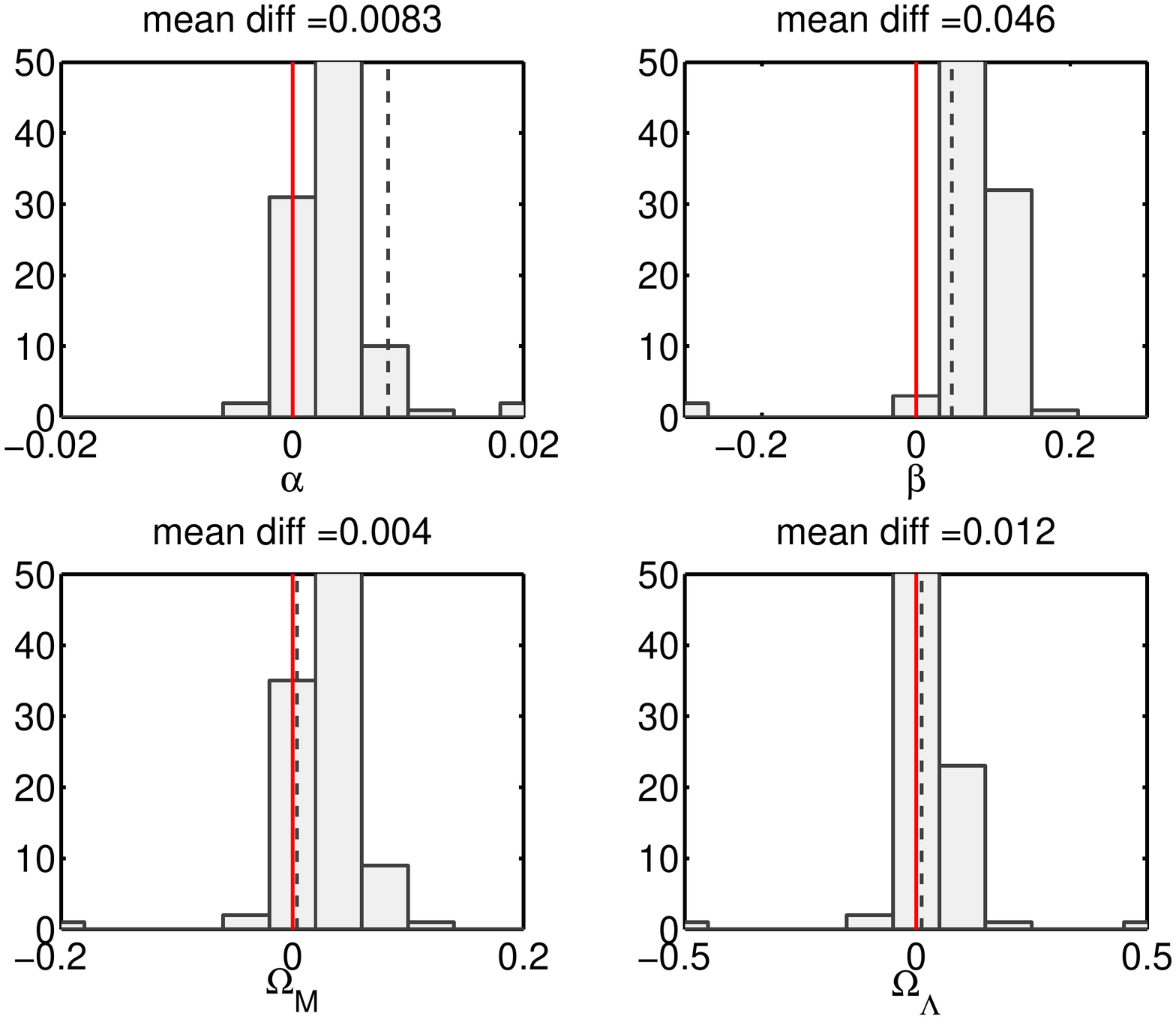}
\includegraphics[width=1.0 \linewidth]{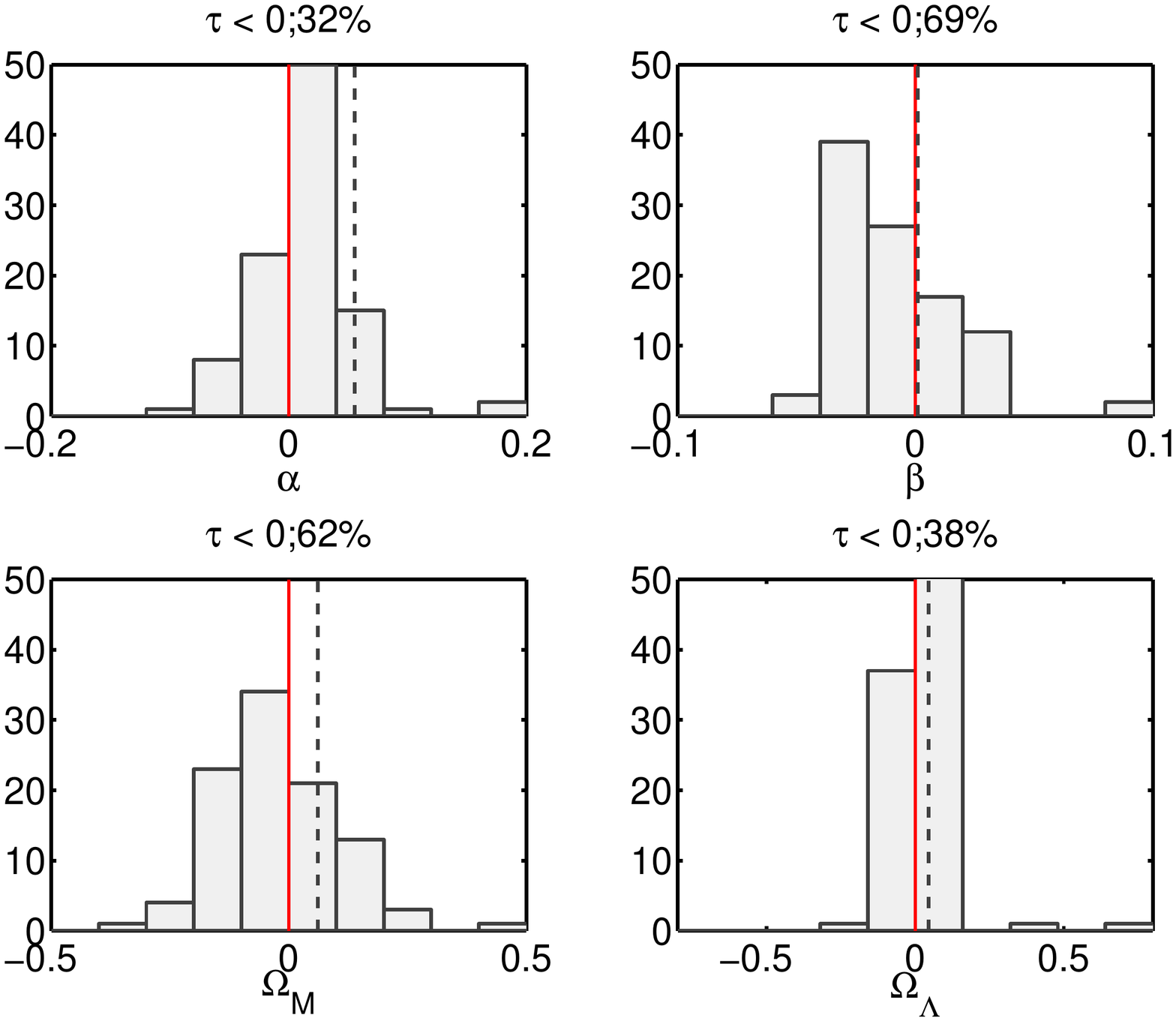} 
\caption{As for Fig.~\ref{fig:lcdm-snls3}, but for the \lcdm~ 
analysis of 100 simulations of the combined `cosmology' data set.}
\label{fig:lcdm-combined}
\end{figure}

The results of these analyses are presented in
Fig.\ref{fig:lcdm-combined}. The top 4 panels show the cosmological
parameter inference results when the SNLS3 SNIa are analysed jointly
with the LowZ, SDSS and HST simulated samples.  As can be seen by
comparing the corresponding panels in Fig. \ref{fig:lcdm-snls3}, the
width of the sampling distribution decreases when the additional low
and high redshift SNIa are included in the analysis.

Increasing the redshift range of the SNIa sample also decreases the
discrepancy between the estimators for the cosmological parameters
given by the two inference methods. From the middle 4 panels of
Fig.~\ref{fig:lcdm-snls3}, the mean differences between the estimators
for $\Omega_{\rm m,0}$ and $\Omega_{\rm \Lambda,0}$ are $0.099$ and
$0.15$ respectively for the SNLS3 sample alone; whereas for the
`cosmology' sample the mean differences between the estimators for
$\Omega_{\rm m,0}$ and $\Omega_{\rm \Lambda,0}$ decrease to $0.004$
and $0.012$ respectively, as can be seen in the middle 4 panels of
Fig.\ref{fig:lcdm-combined}.

\subsection{$w$CDM analysis of combined `cosmological' sample}
\label{sec:simresultswcdm}

As well as the standard \lcdm~ model of the Universe, many theories of
dark energy have been put forward which have a dark energy equation of
state such that $w\ne -1$ (for some examples of comprehensive reviews
see \citealt{LAST2010,FriemanTurner2008,PeeblesRatra2003}). The dark
energy equation of state, $w$ is highly degenerate with curvature, a
degeneracy which cannot be broken with a geometric probe alone such as
the SNIa.  Hence we perform the cosmological parameter inference
within the $w$CDM model in the context of a flat Universe for which
$\Omega_{k,0} \equiv 0$.

\begin{figure}
\centering
\includegraphics[width=1.0 \linewidth]{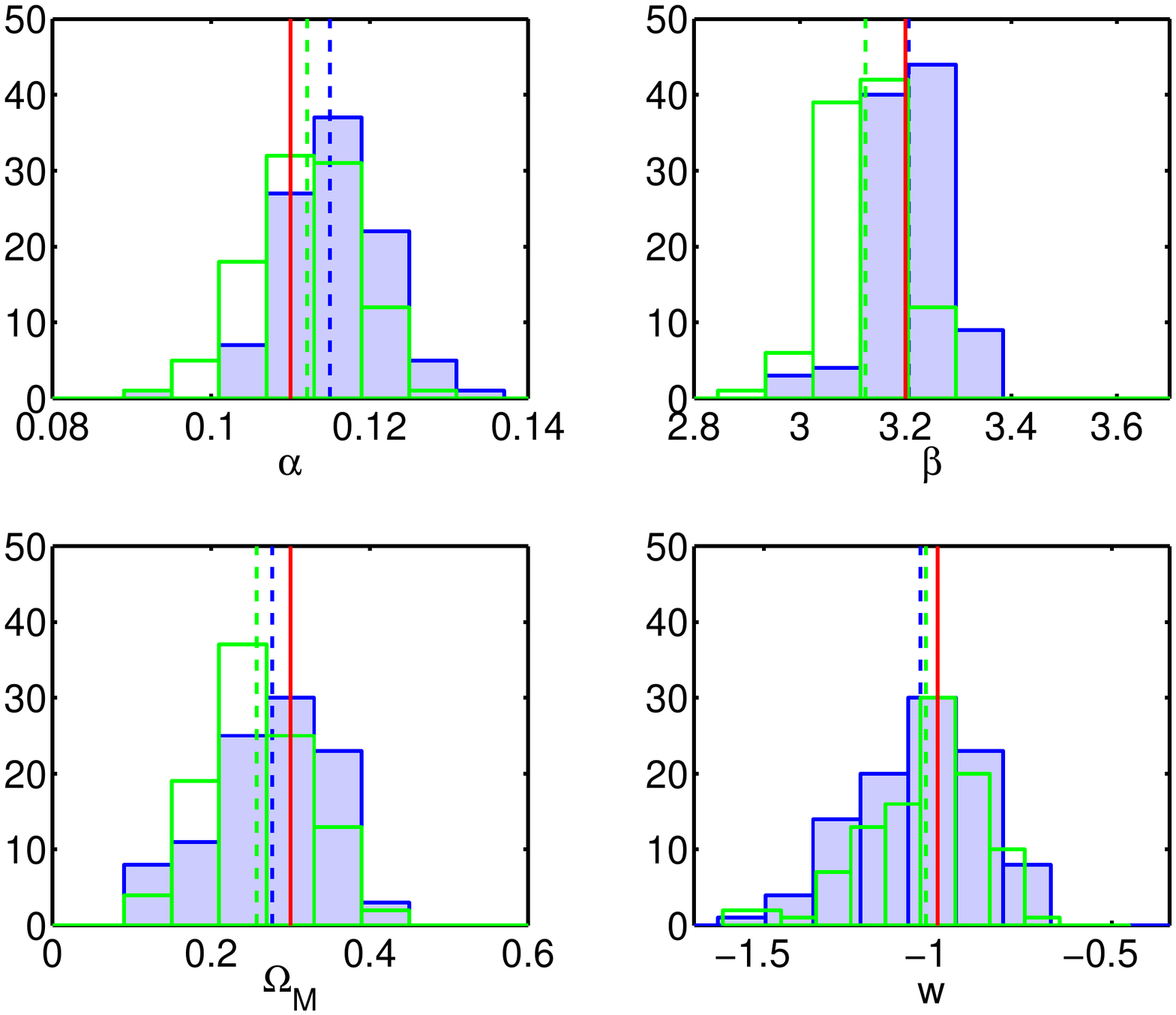}
\includegraphics[width=1.0 \linewidth]{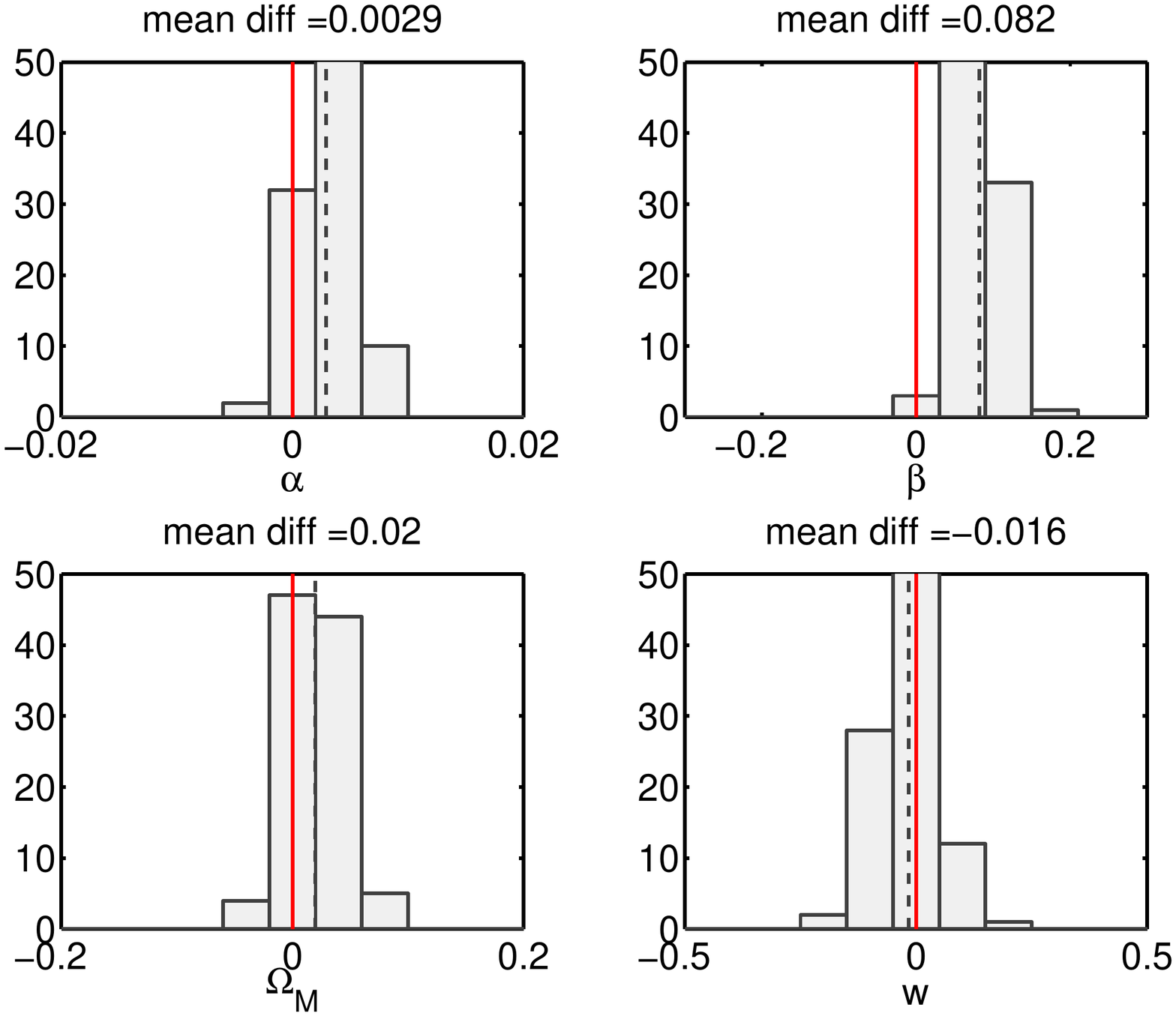}
\includegraphics[width=1.0 \linewidth]{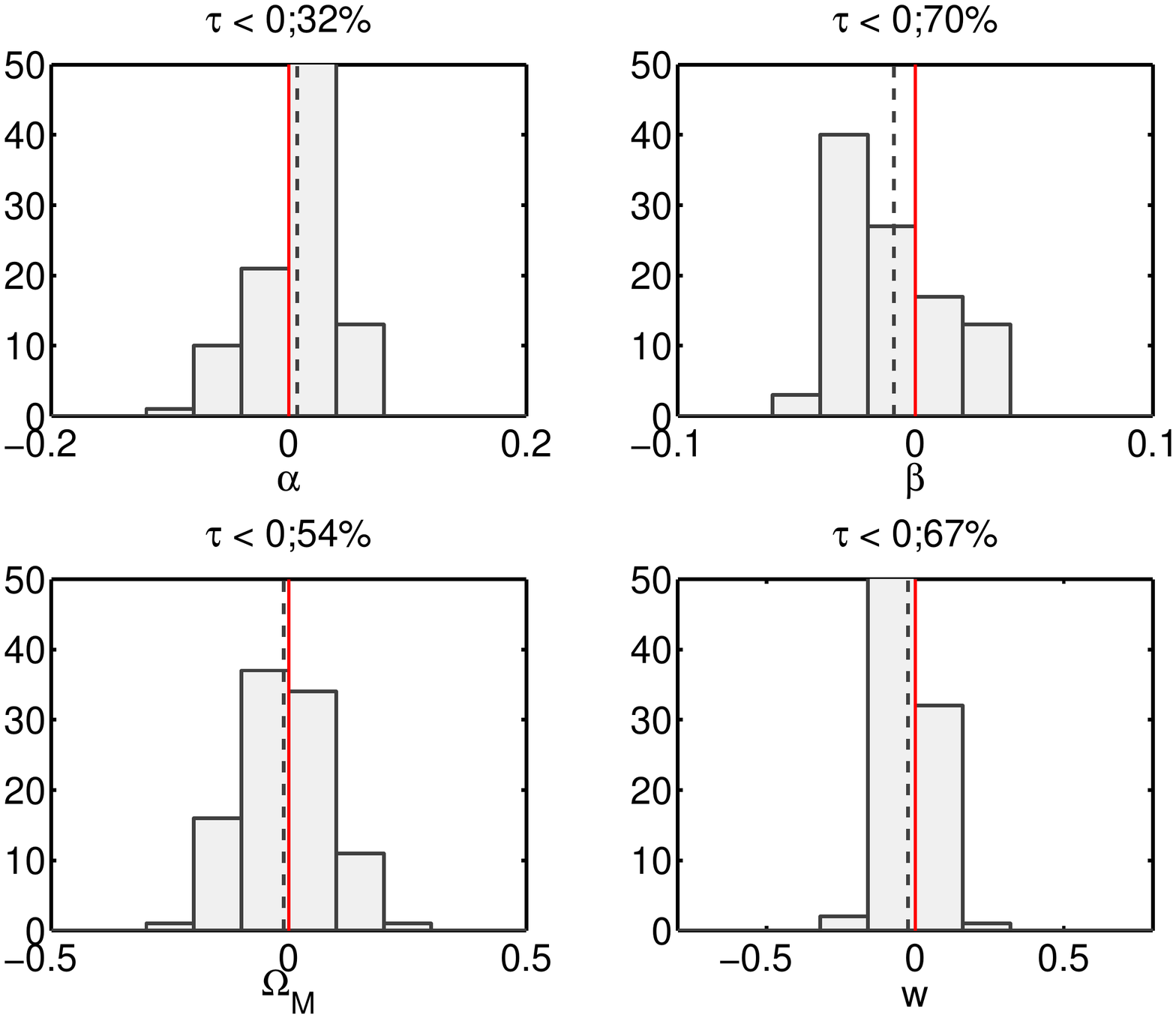} 
\caption{As for Fig.~\ref{fig:lcdm-snls3}, but for the $w$CDM 
analysis of 100 simulations of the combined `cosmology' data set.}
\label{fig:wcdm-combined}
\end{figure}

The SNLS3 sample alone do not cover a broad enough redshift range to
give meaningful constraints on $w$, hence we only present a
cosmological parameter inference for the $w$CDM model using the
combined `cosmology' samples.  The results of the $w$CDM analysis for
the simulated `cosmological' SNIa data sets for the BHM and $\chi^2$
methods are presented in Fig.\ref{fig:wcdm-combined} and
Table~\ref{tab:wcdm-combined}.

\begin{table}
\centering
\begin{tabular}{lrr}
\hline\hline 
                    &\multicolumn{2}{c}{$w$CDM} \\
                    &\multicolumn{2}{c}{Combined sample} \\
Parameter           &  BHM    & $\chi^2$-method \\ \hline
$\alpha$            &$0.0050$ &$0.0021$         \\
$\beta$             &$0.0063$ &$-0.0076$        \\   
$\Omega_{\rm m,0}$  &$-0.023$ &$-0.043$         \\
$w$                 &$-0.050$ &$-0.034$         \\
\hline \hline
\end{tabular}
\caption{Bias on the estimators of the parameters of interest for the $w$CDM model, obtained
  from the two cosmological inference methods applied to the 100 simulated sets of `cosmological' samples.}
\label{tab:wcdm-combined}
\end{table}

%
%

\section{Results from real data}
\label{sec:realresults}

We applied both parameter inference methods to the real SNLS3 single
survey data set, and the combined `cosmological' data set described in
section \ref{sec:realdata}, for the \lcdm~ model and $w$CDM model. The
data sets supplied by the SNLS team are a linearly transformed version
of the \salt{} parameters \citep{GuySullivan2010}, hence it is not
meaningful to compare the transformed SNIa global parameters
$\alpha,\beta$ with the corresponding parameters in our standard
\salt{}.

\begin{figure} \centering
\includegraphics[width=0.8\linewidth]{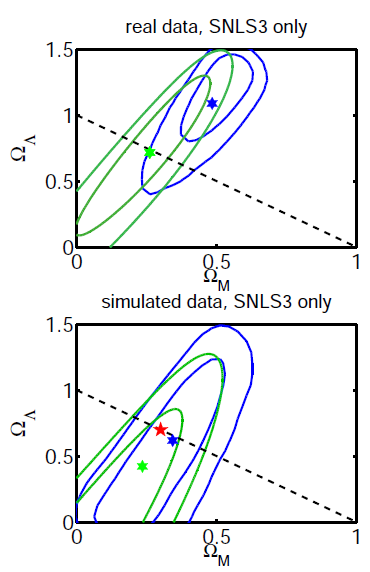}
\caption{Contour plots for \lcdm~ analysis of real and simulated SNLS3 data 
sets. Contours enclose $68.3 \%$ and $95.4
  \%$ of the posterior mass or likelihood for the BHM (blue) and
  $\chi^2$ (green) methods. Blue and green stars show the locations of
  the estimators for the BHM and $\chi^2$ methods respectively.  The
  upper plot shows the analysis of the real data, and the lower plot
  the simulated data. For the analysis of the simulated data, a red star
  indicates the location of the true model input parameters.}
\label{fig:lcdm_contours_snls3}
\end{figure}

\subsection{\lcdm~ analysis of SNLS3 sample}
We present the cosmological parameter inference for the \lcdm~ model
using only SNLS3 data in Fig.~\ref{fig:lcdm_contours_snls3}. The upper
plot in shows the $68.3\%$ and $95.4\%$ contours for the real SNLS3
only data set.  Contours from the BHM are shown in blue, and contours
from the $\chi^2$ method in green.  The estimators for the parameters
for both methods are the expectation values of the 1-D marginalised
posteriors (BHM) or likelihoods ($\chi^2$), the estimators for the BHM
are indicated with a blue star, and the estimators for the $\chi^2$
with a green star. Of note is the discrepancy of $~2\sigma$ in the
inference of $\Omega_{\rm m,0}$ and $\Omega_{\rm \Lambda,0}$,
corresponding to a difference of $~0.2$ units of $\Omega_{\rm m,0}$. A
discrepancy of this magnitude does fall within the expected range of
mean differences (see Fig.\ref{fig:lcdm-snls3}) although it does lie
towards the tail of the mean-difference distribution.  The lower panel
of Fig.~\ref{fig:lcdm_contours_snls3} shows an example analysis of a
corresponding simulated SNLS3 only data set. Additionally, this lower
plot includes a red star to indicate the location of the model input
parameters from which the simulated data set was generated.

\subsection{\lcdm~ analysis of combined `cosmological' sample}
\begin{figure} \centering
\includegraphics[width=0.8\linewidth]{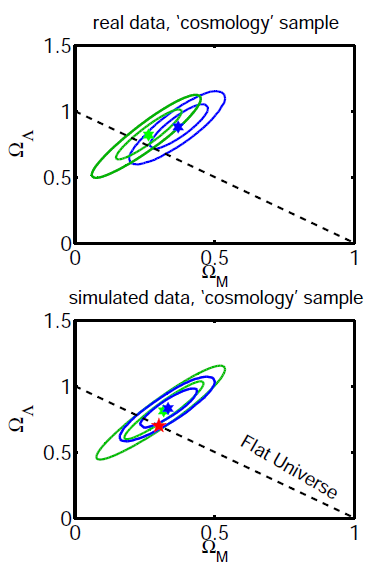}
\caption{As for Fig.~\ref{fig:lcdm_contours_combined}, but for
\lcdm~ analysis of real and simulated combined `cosmological' data sets.} 
\label{fig:lcdm_contours_combined}
\end{figure}

Including the additional high and low redshift SNIa that make up the
`cosmological' sample significantly reduces the discrepancy between
estimators for $\Omega_{\rm m,0}$ and $\Omega_{\rm \Lambda,0}$
obtained using the two different inference methods, as shown in
Fig.~\ref{fig:lcdm_contours_combined}. As expected, for the real data,
increasing the redshift range of the data set greatly enhances the
ability of the SNIa to constrain $\Omega_{\rm m,0}$ and $\Omega_{\rm
  \Lambda,0}$.

Similarly, for the simulated data, including the additional high and
low redshift SNIa decreases the discrepancy between the estimators for
the two methods, and increases the constraining power of the
SNIa. Interestingly, for this particular realization of the simulated
data, both methods give almost the same values for $\Omega_{\rm m,0}$
and $\Omega_{\rm \Lambda,0}$, which are yet $~1\sigma$ distant from
the true values.  There is scope for further investigation as to when
the two methods converge, under what circumstances each method is more
accurate, and in what limit the BHM reduces to the $\chi^2$
approximation for parameter inference.

\subsection{$w$CDM analysis of combined `cosmological' sample}

\begin{figure} \centering
\includegraphics[width=0.99\linewidth]{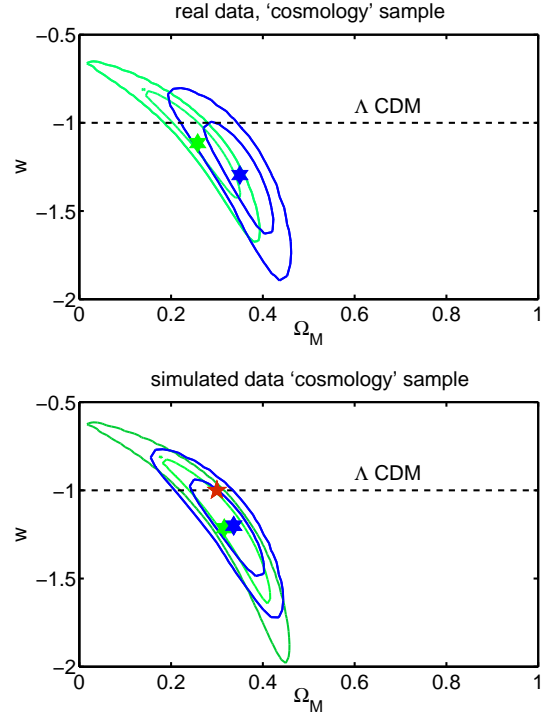}
\caption{As for Fig.~\ref{fig:lcdm_contours_combined}, but for
$w$CDM analysis of real and simulated combined `cosmological' data sets.}
\label{fig:wcdm_contours}
\end{figure}

In Fig.~\ref{fig:wcdm_contours} we present the analysis of the real
`cosmology' data set for the flat $w$CDM model (upper plot), and
compare it with a corresponding simulated data set (lower panel).  The
same simulated data set is used in the lower panels of both
Fig.~\ref{fig:lcdm_contours_combined} and \ref{fig:wcdm_contours}.  The
discrepancy of $~0.2$ units of $w$ between the inferred values for the
dark energy equation of state recovered for the real data set by the
two inference methods is just within the expected range for the mean
difference, as shown in Fig.\ref{fig:wcdm-combined}. As
previously, this particular realisation of simulated data gives a very
small discrepancy between the value of the estimators for the two
methods.

\section{Conclusions}
\label{sec:conc}

Through the analysis of realistic simulated SNe data sets, we have
established that the BHM and $\chi^2$ cosmological parameter inference
methods give different but comparable results when applied to the same
data.  Both methods suffer from small biasses in the recovery of
cosmological parameters and the discrepancy between the two methods is
greatest when only a single survey such as SNLS3 is used.  In this
case, we find that the BHM gives slightly less biassed results,
particularly on the value of $\Omega_{\rm m,0}$. Moreover, the biasses
on $\Omega_{\rm m,0}$ are in opposite directions for the two methods,
which an result in a $\sim 2\sigma$ discrepancy in its estimated value
for any given realisation of the SNLS3-type data. Indeed, we find this
to be the case for the real SNLS3 data set. The discrepancy between
the methods diminishes, however, as the redshift range of the data set
increases. As more higher and lower redshift SNIa are added to the
sample, the two methods begin to converge on their estimate for the
cosmological parameters of interest.

We note that we have investigated the difference between the parameter
inferences obtained for two methods for a given pre-selected,
pre-corrected data set.  One cannot over-emphasise, however, the
importance of the preceding selection cuts and Malmquist bias
corrections for obtaining accurate estimates of the underlying
cosmological parameters. Current SNIa data sets have been
spectroscopically selected, which ensures a high degree of purity of
sample, but future large SN scale surveys such as the Dark Energy
Survey will rely heavily on photometric classification
\citep{BernsteinKessler2012}. For a discussion of how photometric
selection methods affect the cosmological parameter inference
independently of how the parameter inference step is performed, see
\cite{SakoBassett2011}.

Finally, we also point out that our investigation has been carried out
specifically in the context of the \salt{} fitted SNIa data. We
believe this analysis to be of use to the community, as the \salt{}
light curve fitter is one of the most widely used (along with MLCS)
and many current surveys have released their SNIa data in this \salt{}
fitted format. We look forward to the future as increasingly
sophisticated light curve fitting techniques
(e.g.\cite{Mandel:2009xr}) and cosmological parameter inference
methods (e.g. \cite{ShafielooKim2012,SeikelClarkson2012}) gradually
supersede current methods, but in the meantime we offer this study
into the comparative performance of two different cosmological
parameter inference techniques that are currently used in conjunction
with the \salt{} light curve fitter.

\section{Acknowledgements}
We thank Rick Kessler and John Marriner for extensive advice and
assistance in simulating data with SNANA. We thank Heather Campbell
for advice on the SDSS Malmquist bias. We are grateful for useful
conversations and comments from Mark Sullivan, Julien Guy, Mat Smith,
Roberto Trotta. NVK acknowledges support from the Swedish Research
Council (contract No. 621-2010-3301). FF is supported by a Research
Fellowship from Trinity Hall, Cambridge.

\bibliographystyle{mn2e}
\bibliography{cosbench}

\clearpage
\onecolumn
\appendix
\section{Form of the BHM likelihood function}

For each SN, we seek the likelihood (\ref{eqn:likelihood}) of the
input data given the parameters of our model, namely
\begin{equation}
\Pr(\hat{m}_{B,i}^\ast,\hat{x}_{1,i},\hat{c}_i,\hat{z}_i|\Cp,\alpha,\beta,
\sigma_\text{int},\widehat{C}_i,\sigma_{z,i}),
\label{eqn:likefunc}
\end{equation}
where, for completeness, we have included the dependence on the
assumed known uncertaintes $\widehat{C}_i$ and $\sigma_{z,i}$.  As
discussed in Section~\ref{sec:cosmoinf:bhm}, we compute this
likelihood by first introducing the hidden variables $M_i$, $x_{1,i}$,
$c_i$ and $z_i$, which are, respectively, the true (unknown) values of
its absolute $B$-band magnitude, stretch and colour corrections, and
redshift; these are then assigned priors and marginalised over to
obtain the likelihood (\ref{eqn:likefunc}).

For the sake of brevity, we denote the (global) model parameters in
(\ref{eqn:likefunc}) that we wish to constrain by $\bphi =
\{\Cp,\alpha,\beta,\sigma_\text{int}\}$ and those assumed known (and
different for each SN) by $\bpsi_i=\{\widehat{C}_i,\sigma_{z,i}\}$.
Thus, introducing the hidden variables $M_i$, $x_{1,i}$, $c_i$ and
$z_i$, the likelihood (\ref{eqn:likefunc}) can be written as
\begin{equation}
\Pr(\hat{m}_{B,i}^\ast,\hat{x}_{1,i},\hat{c}_i,\hat{z}_i|\bphi,\bpsi) 
= \int dM_i\,dx_{1,i}\,dc_i\,dz_i\,\Pr(\hat{m}_{B,i}^\ast,\hat{x}_{1,i},\hat{c}_i,\hat{z}_i|\bphi,\bpsi,
M_i,x_{1,i},c_i,z_i) \Pr(M_i,x_i,c_i,z_i|\sigma_\text{int}).
\end{equation}
Assuming that the measured redshift $\hat{z}_i$ is independent of 
$\hat{m}_{B,i}^\ast$, $\hat{x}_{1,i}$ and $\hat{c}_i$, and, similarly, that
the true redshift $z_i$ is independent of $M_i$, $x_i$, $c_i$, one may
write
\begin{equation}
\Pr(\hat{m}_{B,i}^\ast,\hat{x}_{1,i},\hat{c}_i,\hat{z}_i|\bphi,\bpsi) 
= \int dz_i\,\Pr(\hat{z}_i|z_i,\sigma_{z,i}) \Pr(z_i)\,\int dM_i\,dx_{1,i}\,dc_i\,
\Pr(\hat{m}_{B,i}^\ast,\hat{x}_{1,i},\hat{c}_i|\bphi,\bpsi,
M_i,x_{1,i},c_i,z_i)
\Pr(M_i,x_{1,i},c_i|\sigma_\text{int}).
\label{eqn:likepart1}
\end{equation}
where $\Pr(z_i)$ is the prior on the true SN redshift, and the prior
$\Pr(M_i,x_{1,i},c_i|\sigma_\text{int})$ can itself be expanded as
\begin{equation}
\Pr(M_i,x_{1,i},c_i|\sigma_\text{int})=\int dM_0\,dx_\ast\,dR_x\,dc_\ast\,dR_c\,
\Pr(M_i,x_{1,i},c_i|M_0,\sigma_\text{int},x_\ast,R_x,c_\ast,R_c)
\Pr(M_0,x_\ast,R_x,c_\ast,R_c|\sigma_\text{int}),
\label{eqn:likepart2}
\end{equation}
in which we have introduced the nuisance hyperparameters $M_0$,
$x_\ast$, $R_x$, $ c_\ast$ and $R_c$ associated with the SN population
and described below. Equations (\ref{eqn:likepart1}) and
(\ref{eqn:likepart2}) form the basis for the Bayesian hierarchical
model of \citet{MarchTrotta2011}.

To proceed further, one first assumes that both of the joint prior
distributions in the integrand of (\ref{eqn:likepart2}) are separable,
as follows:
\begin{eqnarray}
\Pr(M_i,x_{1,i},c_i|M_0,\sigma_\text{int},x_\ast,R_\ast,c_\ast,R_\ast)
& = & \Pr(M_i|M_0,\sigma_\text{int})\Pr(x_{1,i}|x_\ast,R_x)\Pr(c_i|c_\ast,R_c), \\
\Pr(M_0,x_\ast,R_\ast,c_\ast,R_\ast|\sigma_\text{int}) & = &  
\Pr(M_0)\Pr(x_\ast)\Pr(R_x)\Pr(c_\ast)\Pr(R_c).
\end{eqnarray}
One then assigns a form for the prior distribution of each of the
hidden parameters $M_i$, $x_{1,i}$, $c_i$, $z_i$ and nuisance
hyperparameters $M_0$, $x_\ast$, $R_x$, $ c_\ast$, $R_c$. These are
taken to be: $\Pr(M_i|M_0,\sigma_\text{int}) = {\cal N}(M_0,\sigma_{\rm
  int}^2)$, $\Pr(x_{1,i}|x_\ast,R_x) = {\cal N}(x_\ast,R_x^2)$,
$\Pr(c_i|c_\ast,R_c) = {\cal N}(c_\ast,R_c^2)$, $\Pr(z_i)=1$,
$\Pr(M_0)={\cal N}(\bar{M}_0,\sigma_{M_0}^2)$, $\Pr(x_\ast)={\cal
  N}(0,\sigma_{x_\ast}^2)$, $\Pr(c_\ast)={\cal
  N}(0,\sigma_{c_\ast}^2)$, $\Pr(R_x)=1/R_x$ and $\Pr(R_c)=1/R_c$,
where one assumes $\bar{M}_0=-19.3$~mag, $\sigma_{M_0}=2.0$~mag,
$\sigma_{x_\ast}=1$ and $\sigma_{c_\ast}=1$. 

The only remaining probability distributions required to evaluate
(\ref{eqn:likepart1}) are $\Pr(\hat{z}_i|z_i,\sigma_{z,i})$ and
$\Pr(\hat{m}_{B,i}^\ast,\hat{x}_{1,i},\hat{c}_i|\bphi,\bpsi,
M_i,x_{1,i},c_i,z_i)$. The former is given simply by
$\Pr(\hat{z}_i|z_i,\sigma_{z,i})={\cal N}(z_i,\sigma_{z,i}^2)$ and the
latter is the multivariate Gaussian
\begin{equation}
\Pr(\hat{m}_{B,i}^\ast,\hat{x}_{1,i},\hat{c}_i|\bphi,\bpsi,
M_i,x_{1,i},c_i,z_i) = |2\pi \widehat{C}_i|^{-1/2} 
\exp\left[-\tfrac{1}{2}(\hat{\bmath{v}}-\bmath{v})^{\rm t}
\widehat{C}_i^{-1} (\hat{\bmath{v}}-\bmath{v}) \right],
\end{equation}
where
$\hat{\bmath{v}}=[\hat{m}_{B,i}^\ast,\hat{x}_{1,i},\hat{c}_i]^{\rm
  t}$, $\bmath{v}=[\mu(z_i,{\cal C})+M_i-\alpha x_{1,i} + \beta c_i,
  x_{1,i}, c_i]^{\rm t}$ and $\mu(z_i,{\cal C})$ is the predicted
distance modulus given by (\ref{eq:dmod1}).

All the necessary integrals in (\ref{eqn:likepart1}) and
(\ref{eqn:likepart2}) are Gaussian, except those over $R_x$ and
$R_c$. Thus, \citet{MarchTrotta2011} integrate analytically to obtain a
final expression for the likelihood (\ref{eqn:likefunc}) in terms of
an integral over only $R_x$ and $R_c$. Since this expression is rather
complicated, and requires the definition of a number of covariance
matrices, we do not reproduce it here.  In any case, the last two
integrations over $R_x$ and $R_c$ cannot be performed analytically,
and so these variables are added to the parameters of interest
$\bphi$ and sampled, in order to marginalise over them
numerically. It is worth noting that, in principle, all the integrals
in (\ref{eqn:likepart1}) and (\ref{eqn:likepart2}) could be performed
numerically by sampling from the full set of hidden parameters $M_i$,
$x_{1,i}$, $c_i$, $z_i$ and nuisance hyper parameters $M_0$, $x_\ast$,
$R_x$, $ c_\ast$, $R_c$ (in addition to the parameters of interest
$\bphi$), and marginalising over them. Although this would increase
somewhat the dimensionality of the space from which to obtain samples,
it would also allow trivially for more realistic priors on the hidden
and nuisance parameters than the simple separable Gaussian forms
assumed above.  We will investigate this possibility in a future
publication.

\label{lastpage}
\end{document}